# Thermalization-Controlled Superconducting Diode-Like Transport in Disordered Molybdenum and NbN Thin Films

**Akhilesh Kumar[1,2], Shivani Soni[1,2], Gomathi Natarajan[1], Ramanathaswamy Pandian[1,2], Thameemul Ansari M[1,2], Pooja Gupta[2,3], Sanjay Rai[2,3], Edward Prabu Amaladass[1,2*]**

[1] *Material Science Group, Indira Gandhi Centre for Atomic Research, Kalpakkam, 603102*
[2] *Homi Bhabha National Institute, Training School Complex, Anushaktinagar, Mumbai, 400094*
[3] *Synchrotrons Utilisation Section, Raja Ramanna Centre for Advanced Technology, Indore, Madhya Pradesh, 452013, India*

* Corresponding Author Email: edward@igcar.gov.in

## Abstract:

Superconducting diode effects are characterized by direction-dependent critical currents in the superconducting state. Here, we investigate diode-like transport in 128-nm-thick Mo thin film deposited on $Si/SiO_2$ substrate by DC magnetron sputtering. The thin film exhibit a superconducting transition temperature $T_C \approx 5.8$ K. Pronounced current-direction asymmetry and hysteresis are observed in current-voltage characteristics measured in a closed-cycle refrigerator (CCR). In contrast, the same thin film exhibits a substantially more symmetric response in a 14T cryogenic system with improved sample thermalization. We attribute this strong dependence on the measurement environment to electrothermal effects arising from current-induced Joule heating and finite thermal relaxation. As the current approaches the superconducting-to-resistive switching condition, the emergence of a resistive component leads to a rapid increase in Joule heating, raising the sample temperature and modifying the switching condition. Conversely, during the resistive-to-superconducting transition, residual heating and finite thermal relaxation reduce the switching current. Since $P=I^2R$ is independent of current polarity, the observed asymmetry is not associated with polarity-dependent Joule heating but with the interplay of thermal relaxation, sweep history, and thermal boundary conditions. The suppression of the asymmetry under improved thermalization, together with similar behavior observed in NbN reference films, indicates that the apparent diode-like response is strongly influenced by electrothermal effects. These results highlight the need to distinguish thermal switching and hysteresis from intrinsic nonreciprocal superconductivity when identifying superconducting diode effects in transport measurements.

**Keywords:** Superconducting Diode effect (SDE), Nonreciprocal transport, Closed-cycle refrigerator, Joule heating.

## Introduction:

Nonreciprocal transport, in which the electrical response depends on the direction of current flow, is a characteristic manifestation of broken symmetries in quantum materials and has attracted considerable interest in condensed-matter physics and materials science [1-3]. A familiar example is the semiconductor p-n junction, where an asymmetric junction potential produces directional charge transport and the conventional diode effect [4, 5]. In quantum materials, nonreciprocity can arise from the interplay between spatial and time-reversal symmetry breaking [1, 3]. For example, in superconducting systems with broken inversion symmetry, an applied magnetic field can produce a direction-dependent electrical response known as magneto-chiral anisotropy [6-8]. Related effects have been observed in noncentrosymmetric oxide interfaces, chiral conductors [9-11], and organic conductors [12]. These systems establish a general principle: microscopic symmetry breaking can be converted into a measurable directional transport response. Superconductivity provides a particularly intriguing setting for nonreciprocity because charge is carried collectively by a phase-coherent condensate. Nonreciprocal transport has been observed in the superconducting fluctuation regime of systems with broken symmetries [13, 14], where the relevant energy scales is associated with the superconducting gap. This raises a more fundamental question: can directional transport persist in the dissipationless supercurrent itself, rather than being restricted to dissipative quasiparticle transport? The answer is provided by the superconducting diode effect (SDE), in which the critical current depends on the direction of supercurrent, $Ic_+ \neq Ic_-$.Unlike a conventional diode, which rectifies dissipative charge transport, the SDE reflects the nonreciprocity of a coherent superconducting condensate. The possibility of directional superconducting transport predates the modern SDE literature. In 2007, Hu *et al.* theoretically proposed an asymmetric Josephson supercurrent in junctions between p-type and n-type superconductors near the superconducting-Mott-insulator transition [15]. In this Josephson diode scenario, a polar Mott-insulating interface generates an asymmetric supercurrent. Experimentally, a distinct form of directional superconducting switching was observed in 2009 in Py/Nb ferromagnet/superconductor bilayers [16], where the switching current depended on current direction and was attributed to asymmetric screening currents and vortex motion [17]. This vortex-diode mechanism differs from intrinsic condensate-based SDEs, but it demonstrates that broken spatial or interfacial symmetry can produce strongly direction-dependent superconducting states. A decisive realization of the intrinsic SDE was reported by Ando *et al.* in 2020 in a noncentrosymmetric [Nb/V/Ta]n super lattice composed of alternating Nb, V, and Ta superconducting layers [13]. In this structure, inversion symmetry is broken along the out-of-plane c-axis. Under an in-plane magnetic field B, a pronounced asymmetry develops between the critical currents for opposite current directions, with the current flowing perpendicular to the field. The diode signal, defined as $\Delta Ic=Ic_{+}-Ic_{-}$, changes sign under reversal of the magnetic field, establishing the characteristic B-odd SDE [13]. Moreover, driving the system with a current between $Ic_{+}$ and $Ic_{-}$ enables superconducting rectification, switching the device between superconducting and resistive states upon reversal of current direction. This experiment established the SDE as an experimentally accessible form of nonreciprocal superconductivity and stimulated rapid development of the field. Several microscopic mechanisms can produce such directional critical currents, and distinguishing them is essential for interpreting an observed SDE. In noncentrosymmetric superconductors with strong spin–orbit coupling, broken inversion symmetry can generate Rashba-type spin–orbit coupling, while an applied magnetic field breaks time-reversal symmetry. Their combined action can shift the energetic of spin-split bands and stabilize Cooper pairs with finite centre-of-mass momentum, producing a helical or Fulde–Ferrell-like superconducting state [4, 18-20]. Because the condensate then possesses a preferred momentum, super currents flowing in opposite directions can experience different pair-breaking conditions. The resulting critical-current asymmetry provides a direct connection between the momentum structure of the superconducting condensate and the SDE. Its magnitude and polarity can consequently depend on magnetic-field orientation, temperature, spin–orbit coupling, disorder, thickness, and crystallographic direction. Nonreciprocity can also originate from collective vortex dynamics rather than from the intrinsic condensate. In a magnetic field, vortices experience a Lorentz force generated by the transport current. Structural asymmetry, interfaces, defects, pinning landscapes, or magnetic layers can make vortex entry, depinning, and motion inequivalent for opposite current directions [4]. Different vortex barriers or depinning currents can therefore produce different switching currents. The early Py/Nb vortex-diode observation provides an important example of this mechanism [16, 17]. Distinguishing such vortex-mediated nonreciprocity from intrinsic condensate SDEs remains a central experimental challenge.

In Josephson systems, directional supercurrent can arise through an asymmetric current-phase relation (CPR) [21]. For a conventional junction, $I(\phi) = I_C \sin\phi$, giving equal positive and negative critical currents. Broken inversion and time-reversal symmetries, together with spin-orbit coupling or magnetic order, can introduce additional harmonics or anomalous phase shifts into the CPR [22]. An asymmetric CPR then produces unequal positive and negative extrema, thereby yielding a Josephson diode effect. In contrast to vortex-mediated mechanisms, this response directly reflects the phase structure of the Josephson coupling and can therefore provide information about the underlying superconducting order. Magnetic heterostructures provide an additional route to nonreciprocal superconductivity by combining magnetic and structural symmetry breaking. Interfaces between superconductors and ferromagnets, antiferromagnets, or magnetic insulators can introduce exchange interactions or magnetic proximity effects [23], while interfacial structural asymmetry can generate spin-orbit coupling [21].The resulting interplay among superconductivity, magnetism, and spin–orbit coupling can modify the momentum structure of Cooper pairing, thereby controlling both the magnitude and polarity of the diode response[22]. Such systems therefore provide a route to externally manipulating nonreciprocal super currents through magnetic configuration [10, 15, 24-26]. An important extension of the field is the emergence of zero-field and B-even SDEs [15, 27]. In the conventional field-induced SDE, structural inversion-symmetry breaking and an external magnetic field provide the relevant symmetry-breaking ingredients, leading in many cases to a diode polarity that reverses under $B \rightarrow -B$. At zero fields, by contrast, the required time-reversal-symmetry breaking must arise internally.

Possible mechanisms include chiral superconducting order, spontaneous breaking of time-reversal symmetry, magnetic proximity, and asymmetric Josephson coupling [25, 28, 29]. In particular, multicomponent superconducting order can support complex relative phases and chiral states that break time-reversal symmetry. Such an order can generate direction-dependent superconducting energetic and, consequently, a finite difference between $Ic_{+}$ and $Ic_{-}$. Zero-field and B even [7] responses, therefore, provide a distinct window into superconducting states in which nonreciprocity is generated internally rather than imposed by an external magnetic field [30].The growing range of SDE mechanisms-including finite-momentum pairing, vortex dynamics, unconventional CPRs, magnetic proximity, and chiral superconductivity highlights both the versatility and the central challenge of nonreciprocal superconductivity. These mechanisms operate at fundamentally different levels: some originate from the momentum structure of the condensate, others from collective vortex motion or Josephson phase dynamics, and still others from magnetic or unconventional superconducting order. Consequently, a critical-current asymmetry alone does not uniquely determine its microscopic origin. Its dependence on magnetic field, field orientation, temperature, thickness, current direction, and magnetic configuration must be examined systematically to distinguish the underlying mechanisms. Superconducting diode transport thus provides a sensitive probe of symmetry breaking in quantum-coherent superconducting systems. The evolution from early vortex-diode phenomena in ferromagnet/superconductor bilayers to field-induced SDEs in noncentrosymmetric superlattices, and subsequently to zero-field, chiral, and magnetic-heterostructure responses, demonstrates the broad range of physical mechanisms that can produce directional super currents. The experimentally accessible quantity $\Delta Ic=I_{C+} - I_{C-}$connects macroscopic transport directly to the symmetry and microscopic structure of the superconducting state. Establishing the origin of this asymmetry by disentangling inversion-symmetry breaking, magnetic interactions, spin-orbit coupling, vortex physics, and superconducting pairing is therefore essential for developing a quantitative understanding of nonreciprocal superconductivity and for exploiting it in superconducting electronics, spintronics, and quantum technologies.

Motivated by these unresolved issues, an important experimental question is whether diode-like superconducting transport can emerge even in a relatively simple, disordered elemental superconductor without deliberately engineered noncentrosymmetric interfaces or magnetic heterostructures. Here, we investigate 128 nm thick superconducting Mo thin films deposited on $Si/SiO_2$ substrates by DC magnetron sputtering. Although bulk α-Mo has a superconducting transition temperature (Tc) of approximately 0.9 K, the films exhibit Tc = 5.8 K, indicating a substantial modification of the superconducting state associated with the film microstructure and disorder. Interestingly, I-V measurements performed in a CCR reveal pronounced current-direction asymmetry resembling a superconducting diode response. However, measurements of the same film in a 14 T cryogenic system with enhanced sample thermal stabilization yield a substantially more symmetric response. The control experiment on NbN thin films also exhibits similar behavior. The recovery of nearly reciprocal transport under improved thermal stabilization highlights the crucial role of thermal control in distinguishing genuine superconducting diode effects from measurement-induced diode-like transport. Thermal simulation using Energy 2D also supports the distinct thermalization of the sample in these two measurement conditions.

## 2. Experimental Method:

Molybdenum (Mo) thin films were deposited on $Si/SiO_2$ substrates using DC magnetron sputtering. Prior to deposition, the substrates were boiled in deionized (DI) water for 3 min and subsequently ultrasonically cleaned in DI water, acetone, and isopropyl alcohol. A Mo target of 99.95% purity from Testbourne, UK was used as the sputtering source. Before introducing argon, the chamber was evacuated to a base pressure of $3 \times 10^{-6}$ mbar. Deposition was then carried out for 15 min at a constant DC power of 50 W under an argon working pressure of $3.2 \times 10^{-3}$ mbar. Surface morphology was examined using a Carl Zeiss Crossbeam 340 SEM operated in in-lens duo mode at 3-5 kV, and cross-sectional SEM imaging was used to determine the film thickness. Grazing incidence X-ray diffraction (GI-XRD) measurements were carried out at beam line (BL-2), RRCAT Indore, to analyze the crystallinity and phase composition of the films. Raman spectroscopy was performed at room temperature using a 532 nm laser (InVia, Renishaw). The superconducting and transport properties were investigated using a four-probe configuration. Temperature-dependent resistance, current-voltage (I-V), and low-temperature magneto-transport measurements were initially performed in a closed-cycle refrigerator (Oxford Instruments, Model RDK-101D). A Keithley 6221 DC and AC source or a Keithley 2400 DC source was used for current biasing, and a Keithley 2182A nanovoltmeter was used to measure the voltage response. The sample temperature was controlled using an Oxford Instruments Mercury iTC temperature controller. The I-V characteristics were measured by sweeping the current from +20 mA to −20 mA with a step size of 1 mA while recording the corresponding voltage. To investigate the transport response under improved thermal stabilization, measurements were also performed in a 14 T cryogenic magneto-transport system (Cryogenic Ltd., UK) offering enhanced cooling efficiency. The same four-probe configuration, current source, and nano-voltmeter were used for these measurements. Temperature-dependent resistance, magnetic-field-dependent transport, and I-V characteristics were measured over the relevant temperature and magnetic-field ranges. The Mo thin films were investigated in both cryogenic systems using the same electrical measurement configuration. NbN thin-film samples were also measured in both systems under identical electrical conditions to compare the two measurement environments.

## 3. Results and Discussion:

### 3.1 Enhanced superconductivity and anomalous nonreciprocal transport

The superconducting properties of the sputtered Molybdenum thin film were first examined through temperature-dependent electrical resistance. The film exhibits a sharp superconducting transition at Tc ≈ 5.8 K, as shown in Figure 1(a), substantially above the bulk transition temperature of Molybdenum (Tc ≈ 0.9 K). This pronounced enhancement of $T_C$ demonstrates that the superconducting state in the sputtered film is modified from that of bulk Molybdenum, reflecting the influence of disorder, microstructural evolution, and strain associated with thin-film growth. An unexpected feature emerges when the superconducting state is probed by DC current-voltage (I-V) measurements. Below Tc, the I-V characteristics show pronounced hysteresis and a clear difference between the positive (I+) and negative (I-) current directions, as shown in Figure 1 (b). We measured the I-V characteristics under applied magnetic fields, as shown in Figure 1(c). The I-V curves retain their hysteretic behavior and current-direction asymmetry under a magnetic field, indicating that the nonlinear transport response persists beyond the zero-field condition. With increasing magnetic field, the critical current $J_C$ decreases systematically, as expected from the progressive suppression of superconducting transport by the applied field.

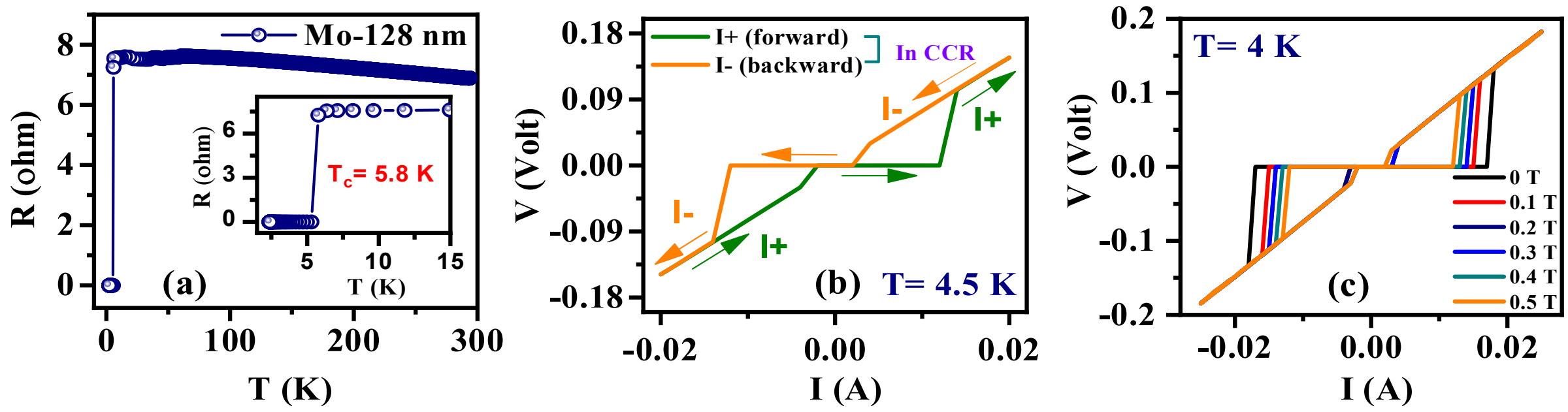

**Figure 1. Superconducting transition and field-dependent nonreciprocal transport in Mo thin films.**(a) Temperature dependence of the resistance, revealing a superconducting transition at Tc ~ 5.8 K.(b) Representative I-V characteristics measured in a closed-cycle refrigerator (CCR) at T =4.5 K, showing pronounced polarity-dependent switching between the I+ and I- branches and a corresponding diode-like asymmetry. (c) I- V characteristics measured in the same CCR at T= 4 K under different applied magnetic fields. The I+ and I- branches remain distinct over the measured magnetic-field range, with the polarity asymmetry persisting and evolving with field, indicating a field-dependent nonreciprocal transport response.

In particular, the switching currents differ between the two current polarities, resulting in a pronounced diode-like, nonreciprocal transport response. The simultaneous observation of hysteresis and current-direction asymmetry indicates that the superconducting transport is nonlinear and may involve weakly coupled or spatially inhomogeneous superconducting regions. The observation is particularly intriguing because the film contains no deliberately engineered magnetic layer, noncentrosymmetric interface, or lithographically defined Josephson junction. This raises an important question: can the intrinsic microstructure of a disordered elemental superconducting film provide the weak-link environment required for such anomalous transport?

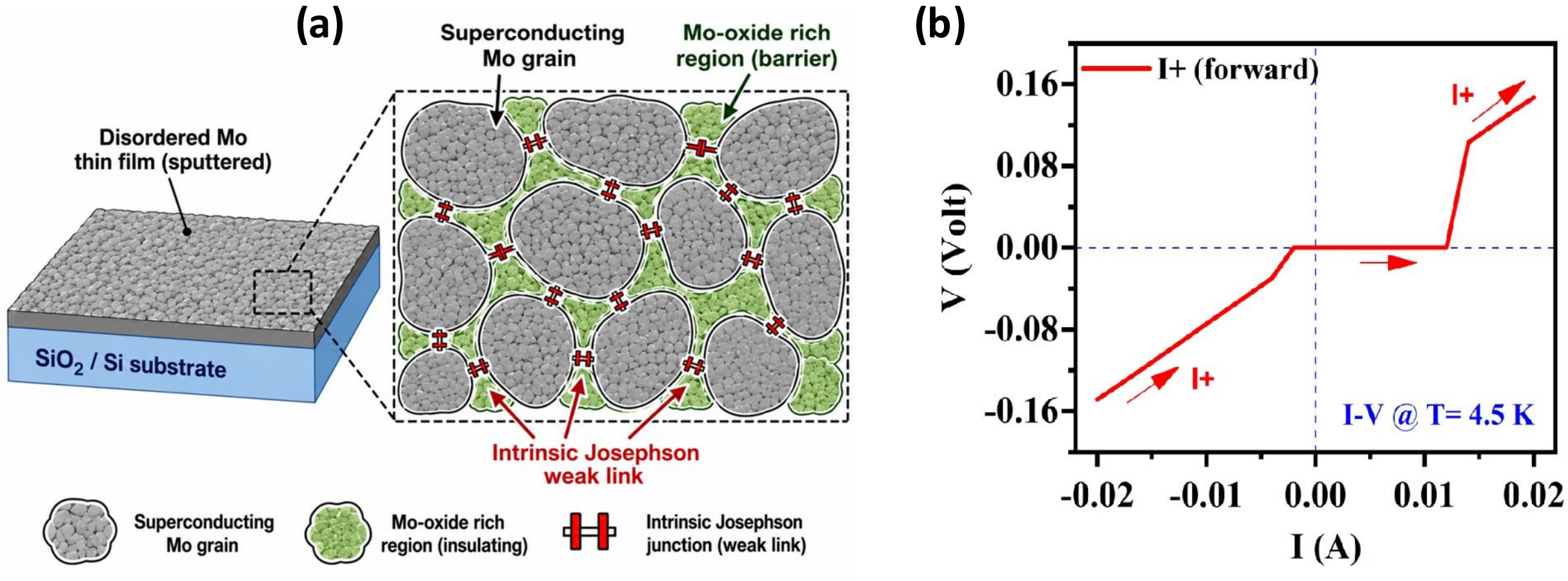

**Figure 2. Granular superconducting microstructure and possible weak-link-mediated diode-like transport.** (a) Schematic illustration of the sputtered Mo thin film, highlighting superconducting Mo grains embedded in a Mo-oxide-rich insulating matrix. The enlarged view illustrates the resulting possible network of weak-link pathways connecting neighboring superconducting grains via oxide-rich regions. (b) Representative I-V characteristic measured at T = 4.5 K, exhibiting distinct I+ and I- branches and pronounced polarity asymmetry. The schematic microstructure and corresponding asymmetric transport highlight the role of intrinsic weak links in the nonreciprocal response of the granular superconducting film.

A plausible mechanism can be proposed based on the chemical nature of Molybdenum. In a thin-film environment, partial oxidation may lead to the coexistence of superconducting Mo grains and Mo-oxide-rich regions. When these phases are distributed over sufficiently small length scales, the oxide regions can act as weak barriers between neighboring superconducting grains, thereby forming naturally occurring Josephson weak links, as shown in Figure 2 (a). Such micro-structural in homogeneity may therefore provide a pathway for Josephson coupling between spatially separated superconducting Mo grains. These weak links may influence the nonlinear current transport and contribute to the observed hysteresis and current-direction asymmetry as shown in Figure 2 (b). Importantly, however, the observation of asymmetric I-V characteristics alone does not establish an intrinsic superconducting diode effect. Nonreciprocal transport can arise through several microscopic mechanisms, and weak-link systems are particularly susceptible to thermal fluctuations and current-induced heating. We therefore first examine whether the Mo film exhibits structural and morphological features that could provide a natural pathway for weak Josephson coupling within the film.

## 3.2 Structural and Morphological Fingerprints of Anomalous Transport

To characterize the structural and microstructural properties of the deposited Mo thin film, we performed Grazing-incidence X-ray diffraction (GI-XRD), Raman spectroscopy and scanning electron microscopy (SEM) measurements. These techniques were used to identify the crystalline phases, probe vibrational modes, and examine the surface morphology, with particular emphasis on oxide formation and structural inhomogeneity. Such features are relevant for assessing the possible presence of weak-link regions and their contribution to the observed anomalous superconducting transport.

### 3.2.1 Grazing-incidence X-ray diffraction (GI-XRD) analysis

The structural characteristics of the sputtered Mo thin film were investigated using grazing-incidence X-ray diffraction (GIXRD). The diffraction pattern reveals the coexistence of crystalline Mo and molybdenum oxide phases, indicating that the deposited film is not a single-phase metallic Mo film. The prominent diffraction peaks are attributed to crystalline Mo, while additional reflections can be indexed to Mo-oxide phases, including $MoO_2$ and $MoO_3$, as shown in Figure 3 (a). The simultaneous presence of metallic Mo and oxide phases provides plausible structural evidence of the formation of a mixed Mo/Mo-oxide structure within the sputtered thin film. The presence of these oxide phases suggests the inadvertent incorporation of oxygen during film deposition or subsequent exposure to the ambient environment, resulting in partial oxidation of the deposited Mo film. The presence of oxide-related diffraction features is particularly interesting in the context of the anomalous superconducting transport discussed in Section 3.1.

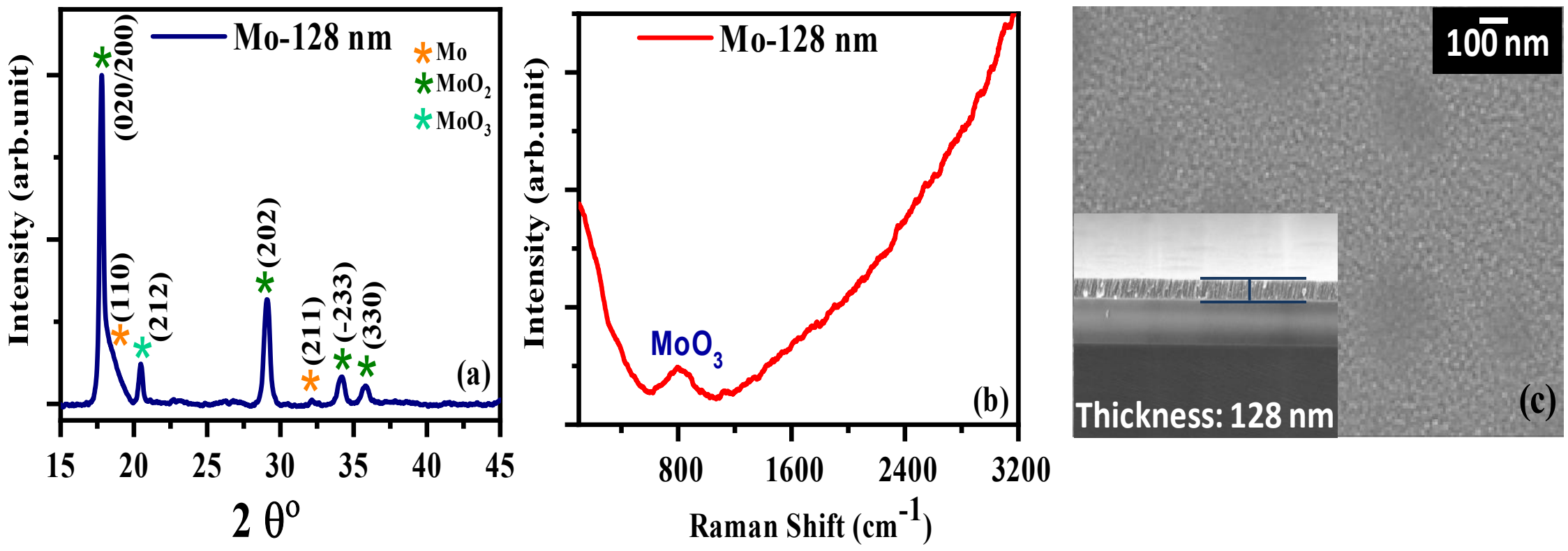

**Figure 3. Structural fingerprints and granular morphology of the sputtered Mo film**.(a) X-ray diffraction pattern revealing reflections from Mo, $MoO_2$, and $MoO_3$, indicating the multiphase nature of the sputtered thin film. (b) Raman spectrum showing a characteristic $MoO_3$ vibrational mode near 821 cm$^{-1}$, providing further evidence of $MoO_3$ in the thin film. (c) SEM image revealing the granular surface morphology of the thin film; in the inset cross-sectional SEM confirms a thickness of thin film approximately 128 nm.

Importantly, the GIXRD results alone do not establish the presence of Josephson weak links or determine the distribution of the Mo and Mo-oxide phases at the nanoscale. However, the clear identification of multiple Mo-oxide phases establishes an important structural feature of the thin film that could support a non-uniform superconducting network. This observation motivates further investigation of the local vibrational modes and surface morphology using Raman spectroscopy and SEM, respectively, to determine whether the oxide formation is accompanied by corresponding morphological variations.

### 3.2.2 Raman spectroscopic analysis

The Raman spectrum of the Mo thin film is shown in Figure 3 (b).A broad hump centered on 821 cm$^{-1}$ is observed, which can be attributed to the characteristic vibrational modes of $MoO_3$ [31]. This Raman feature further supports the presence of $MoO_3$ in the thin film and is consistent with the Mo-oxide phases identified by GI-XRD.

### 3.2.3 SEM morphological analysis

The surface morphology of the sputtered Mo thin film was examined using SEM, as shown in Figure 3 (c). The SEM image reveals a granular morphology with clearly visible grains and grain boundaries. The grains are closely packed across the surface, indicating a uniform granular structure. The cross-sectional SEM image shown in the inset gives a film thickness of approximately 128 nm. The film appears continuous across its thickness.

Overall, the GI-XRD, Raman, and SEM measurements provide consistent evidence of structural and morphological inhomogeneity in the Mo thin film. GI-XRD reveals crystalline Mo along with $MoO_2$ and $MoO_3$ phases; Raman spectroscopy confirms the presence of $MoO_3$-related vibrational features; and SEM shows a continuous film with patchy surface morphology. The coexistence of metallic Mo and oxide phases provides a plausible structural picture in which superconducting Mo regions are locally separated by oxide-rich regions. Such regions can reduce the coupling between neighboring superconducting regions and may provide naturally occurring weak-link pathways within the film. This micro structural picture, therefore, offers a possible connection between the film structure and the anomalous nonlinear and nonreciprocal transport observed experimentally.

### 3.2.4 Experimental approach to distinguish weak-link and heating effects

The Mo results suggest two possible origins of the observed nonreciprocal transport: (i) possible oxide-mediated weak-link regions associated with the coexistence of Mo and Mo-oxide phases, and (ii) Joule heating during the current sweep. To distinguish between these possibilities, an NbN thin film was investigated as a reference superconducting system. NbN is highly stable against oxidation under the present conditions and, therefore, is not expected to develop oxide-mediated Josephson weak links similar to those proposed in the Mo film. It thus provides a suitable system for examining whether nonreciprocal transport can arise from Joule heating in the absence of such naturally formed weak links.

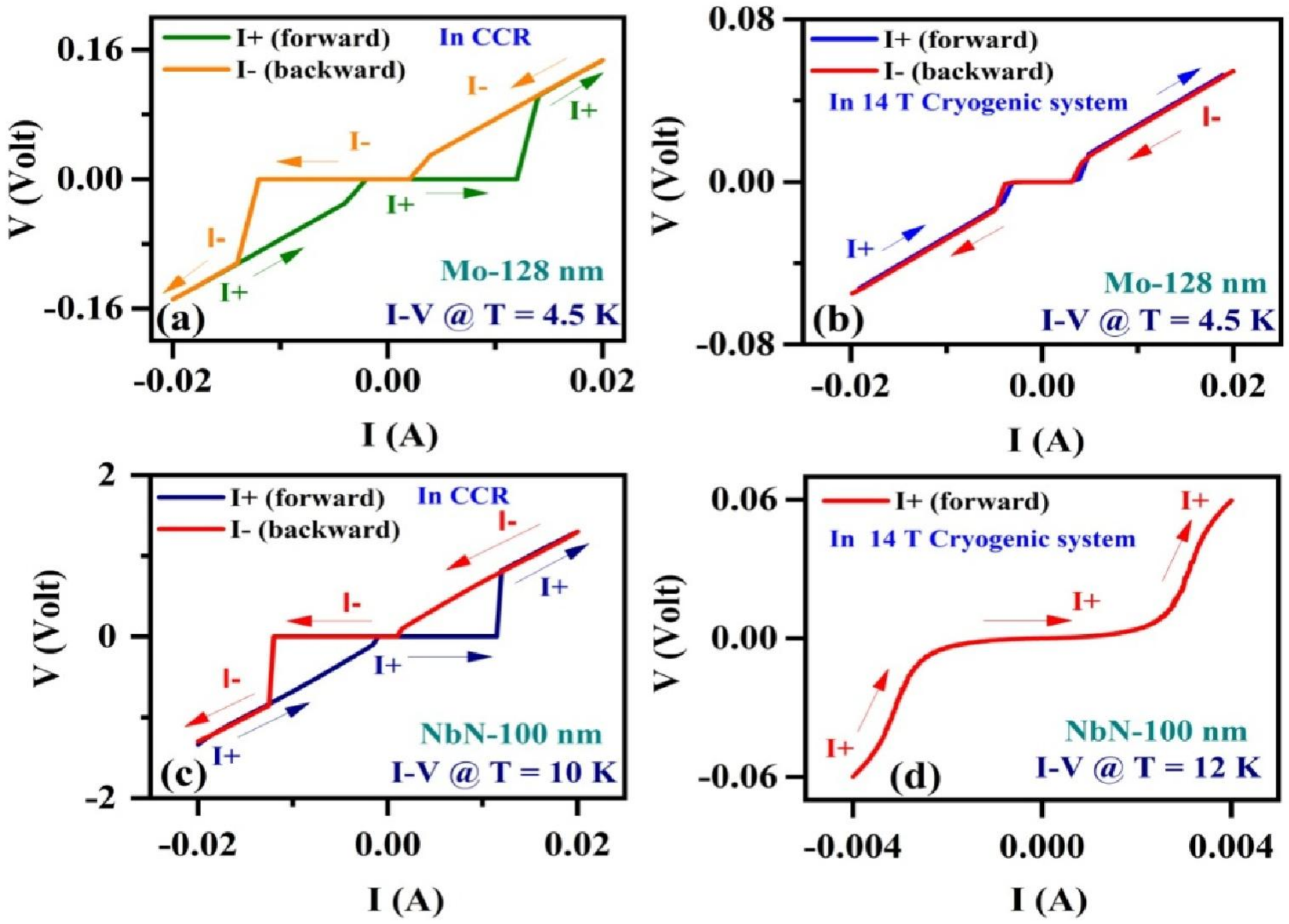

**Figure 4. Comparative transport response of Mo and NbN superconducting thin films**. (a, b) I-V characteristics of the 128 nm thick Mo thin film at T = 4.5 K measured in a closed-cycle refrigerator (CCR) and a 14 T cryogenic system. The pronounced separation of the I+ and I- branches in the CCR contrasts with the substantially more symmetric response obtained in the 14 T cryogenic system. (c) I-V characteristics of a 100 nm thick NbN thin film at T =10 K measured in the CCR, exhibiting a pronounced polarity asymmetry between the forward and reverse current branches. (d) I-V characteristic of the 100 nm thick NbN thin film at T =12 K measured in the 14 T cryogenic system, exhibiting a nearly symmetric response.

For this comparison, both the Mo and NbN thin films were measured in two low-temperature cryogenic systems with different cooling mechanism. The first was a closed-cycle refrigerator (CCR), here the sample is directly mounted on the cold finger and the heat removal from the sample occurs mainly through thermal conduction. The second was a 14 T cryogenic system, which provides more efficient cooling by both thermal conduction through the sample holder and thermal convection by the helium vapor-mediated heat transfer. Comparing the NbN transport response in these two systems allows the effect of cooling efficiency on the current-induced response to be examined directly. This provides a controlled way to assess the contribution of Joule heating and to distinguish it from the weak-link mechanism proposed for the Mo film. This approach provides two complementary tests: the NbN measurements examine whether nonreciprocal transport can occur without the oxide-mediated weak-link structure proposed for Mo. In contrast, the comparison between the CCR and 14 T Cryogenic system examines how the response changes with cooling efficiency. Figure 4 compares the I-V characteristics of the Mo and NbN films measured under the two cryogenic conditions. For the 128 nm thick Mo thin film measured in CCR at T = 4.5 K, as shown in Figure 4 (a), the I-V characteristics show pronounced hysteresis, with distinctly different switching currents in the two current directions. Remarkably, for the same film measured in the 14 T cryogenic system, the hysteretic I-V behavior persists, but the pronounced current-direction asymmetry is largely suppressed, resulting in a nearly symmetric I-V response, as shown in Figure 4 (b). For quantitative analysis the diode-like shift in the I-V characteristic is quantified by the parameter ($\eta$) calculated using equation.1 given below.

$$\eta = \frac{|I_{C+}| - |I_{C-}|}{|I_{C+}| + |I_{C-}|} \qquad \text{(Equation. 1)}$$

Here $I_{C-}$ is the switching current corresponding to the transition from the normal state to the superconducting state and $I_{C+}$ is the switching current corresponding to the transition from the superconducting state to the normal state during a negative to positive current sweep.

For the 128 nm thick Mo film $\eta$ decreases from 0.71 in the CCR to 0.16 in the 14 T cryogenic system. A similar evolution is observed in the 100 nm NbN thin film, as shown in Figure 4 (c). At T = 10 K, the NbN film exhibits hysteretic and nonreciprocal I-V characteristics in the CCR, despite the absence of the oxide-mediated weak-link structure proposed for the Mo film. When the same NbN thin film is measured in the 14 T cryogenic system at T = 12 K, the current-direction asymmetry is strongly suppressed, and the response becomes nearly symmetric for positive and negative current sweeps, with no appreciable hysteresis or current-direction asymmetry, as shown in Figure 4 (d).The deduced $\eta$ for NbN thin film decreases from 0.84 in the CCR to zero in the 14 T cryogenic system. Thus, $\eta$ decreases by ~ 77.5% (4.4-fold reduction) in Mo thin film, whereas the diode-like response vanishes entirely in NbN thin film under the 14 T cryogenic measurement conditions.

The observed diode-like response can be understood in terms of the competition between Joule heating and the measurement systems effective heat removal and thermalization conditions. In the CCR system, where cooling of the sample is primarily mediated by conduction, the CCR (Oxford Instruments RDK-101D) provides a nominal cooling capacity of approximately 1-1.5 W at 4.2 K. Despite this substantial nominal cooling capacity, the effective thermal conductance between the thin film and the cold stage determines how rapidly heat generated in the thin film is removed. During the superconducting state to normal state transition, the superconducting thin film initially dissipates very little power because its resistance is negligibly small. As the applied current approaches, the switching condition and a resistive component develop, the dissipated power $P = I^2R$ increases rapidly. Conversely, during the resistive/normal state-to-superconducting transition, residual heating and finite thermal relaxation reduce the switching current. This self-heating can unequally raise the local film temperature and modify the switching condition. Such electrothermal feedback is known to produce abrupt current-driven transitions and hysteresis in superconducting films [32, 33]. It is therefore important to distinguish the intrinsic switching current density (Jsw) from the experimentally observed switching current density (Jsw). The transition from the superconducting state to the resistive/normal state identifies a switching current under the particular thermal boundary conditions and current-sweep protocol. Still, it cannot by itself be regarded as a direct measurement of the equilibrium intrinsic Jc when appreciable self-heating is present. Likewise, the return transition from the resistive state toward the superconducting state can occur at a different current because the sample may remain thermally elevated. Therefore, it always yields a lower switching current than the superconducting-to-resistive transition branch. This difference between switching and retrapping conditions provides a natural origin for the observed hysteresis. The polarity dependence requires additional care. Since $P = I^2R$ is independent of the sign of current, Joule heating alone does not produce a larger power for positive current than for negative current at equal |I| and R. The observed separation of the positive and negative branches can instead result from the finite thermal relaxation time, the fixed sweep history used in the measurement, and any polarity-dependent electrical or thermal asymmetry of the measurement circuit. Thus, the observed current-direction asymmetry is more appropriately described as thermalization-controlled, diode-like transport rather than as direct evidence that the intrinsic critical current differs between the two current polarities. This response changes markedly when the sample is measured in a configuration that provides more effective thermal stabilization. In the 14 T cryogenic system, the nominal cooling capacity at 4.2 K is approximately 0.2-0.5 W, lower than that of the CCR; however, the present configuration provides more effective thermal coupling to the sample through the combined heat-removal pathways described in the experimental section. This enhanced thermalization suppresses the temperature excursion associated with current-induced heating and correspondingly reduces the apparent current-direction asymmetry. The schematic of this different cooling and thermalization is shown in Figure 5 (a-b).

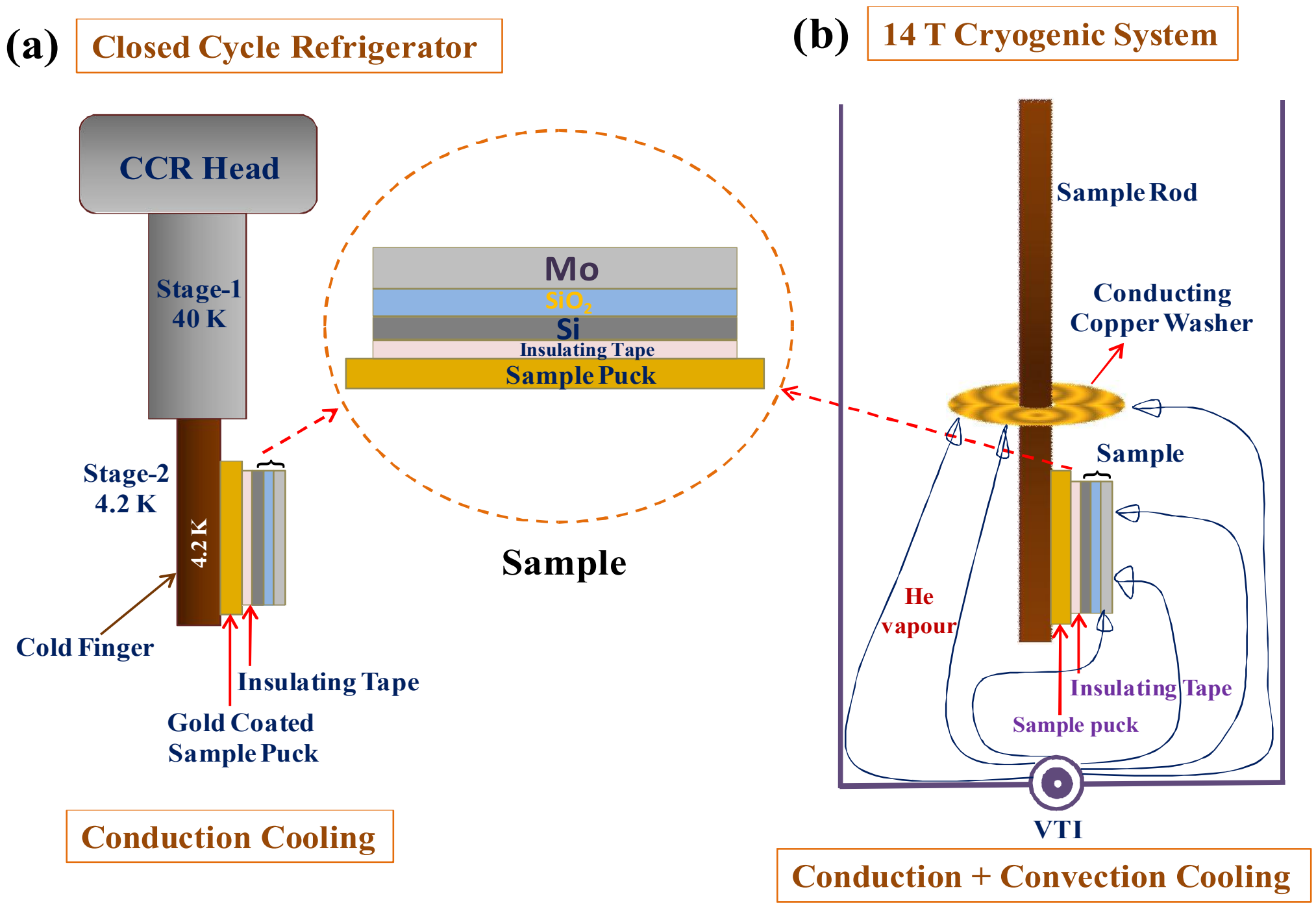

**Figure 5. Schematic illustration of sample mounting and thermalization pathways in two cryogenic environments.**(a)In the closed-cycle refrigerator (CCR), the Mo thin film deposited on a $Si/SiO_2$ substrate is mounted on a gold-coated sample holder, electrically isolated from the holder by an insulating layer and thermally coupled to the CCR cold finger maintained at 4.2 K. (b) In the 14 T cryogenic system, the sample is mounted on a gold-coated sample puck, separated from the mount by an insulating layer and thermally coupled to the copper cold finger. In addition to conductive heat transfer through the sample mount, the sample is also exposed to convective cooling from the surrounding helium vapor. The schematic highlights the distinct heat-removal pathways in the two configurations, illustrating how the additional convective cooling pathway can improve sample thermalization and mitigate Joule-heating-induced temperature excursions during electrical transport measurements.

### 3.2.5. Thermal Simulation:

To investigate the thermalization behaviour, simulations were performed using the free Energy 2D simulation package [34].The sample was represented by multilayer stacks of gold-coated sample holder/insulating tape/$Si/SiO_2/Mo$ and gold-coated sample puck/insulating tape/$Si/SiO_2/NbN$, with the corresponding dimensions, thermal conductivities, specific heat capacities, and densities assigned to each layer. The gold-coated copper sample holder measured 4 cm × 4 cm and was 1 mm thick, while the remaining layers measured 6 mm (length) × 4 mm (width). The thicknesses of the insulating tape, Si substrate, $SiO_2$ layer, and Mo film were 200µm, 500 µm, 300 nm, and 128 nm, respectively. Room-temperature values of thermal conductivity, specific heat capacity, and density were assigned to each material as follows: copper (401 W $m^{-1}$ °$C^{-1}$, 385 J $kg^{-1}$ °$C^{-1}$, 8933 kg $m^{-3}$); insulating tape (0.12 W $m^{-1}$ °$C^{-1}$, 1090 J $kg^{-1}$ °$C^{-1}$, 1420 kg $m^{-3}$); silicon (130 W $m^{-1}$ °$C^{-1}$, 700 J $kg^{-1}$ °$C^{-1}$, 2329 kg $m^{-3}$); $SiO_2$ (1.4 W $m^{-1}$ °$C^{-1}$, 730 J $kg^{-1}$ °$C^{-1}$, 2200 kg $m^{-3}$); and Mo (138 W $m^{-1}$ °$C^{-1}$, 251 J $kg^{-1}$ °$C^{-1}$, 10,240 kg $m^{-3}$). For NbN, the corresponding values were 3.8 W $m^{-1}$ °$C^{-1}$, 365 J $kg^{-1}$ °$C^{-1}$, and 7300 kg $m^{-3}$, respectively [35-37]. These parameters were used to compare heat-transfer pathways in the two sample configurations. An initial temperature gradient was established between the top and bottom surfaces of the multilayer structure. The system was then allowed to evolve toward thermal equilibrium under two conditions: (i) conduction only heat transfer and (ii) combined conduction and convection. In the conduction-only case, heat transfer occurs through the solid layers, whereas in the second case, convective heat transfer was additionally included. The temperature distribution was monitored at thermalization times of 30 s, 1 min, and 4 min as shown in Figure 6 (a-c). The simulations were used to compare the thermal relaxation dynamics of the two sample structures and the two heat-transfer conditions. No Joule-heating power was used as a source term in these simulations; the imposed temperature gradient was used solely to examine the subsequent thermal equilibration. The simulations show faster thermal equilibration when convection is included. This provides a qualitative representation of enhanced heat removal and is consistent with the faster thermalization observed experimentally in the 14 T configurations. The simulation does not explicitly model helium-vapor flow or the detailed cryogenic boundary conditions of the experimental system. These simulations therefore support the conclusion that the effective thermal boundary condition, rather than nominal cryostat cooling capacity alone, is an important parameter governing the observed transport response.

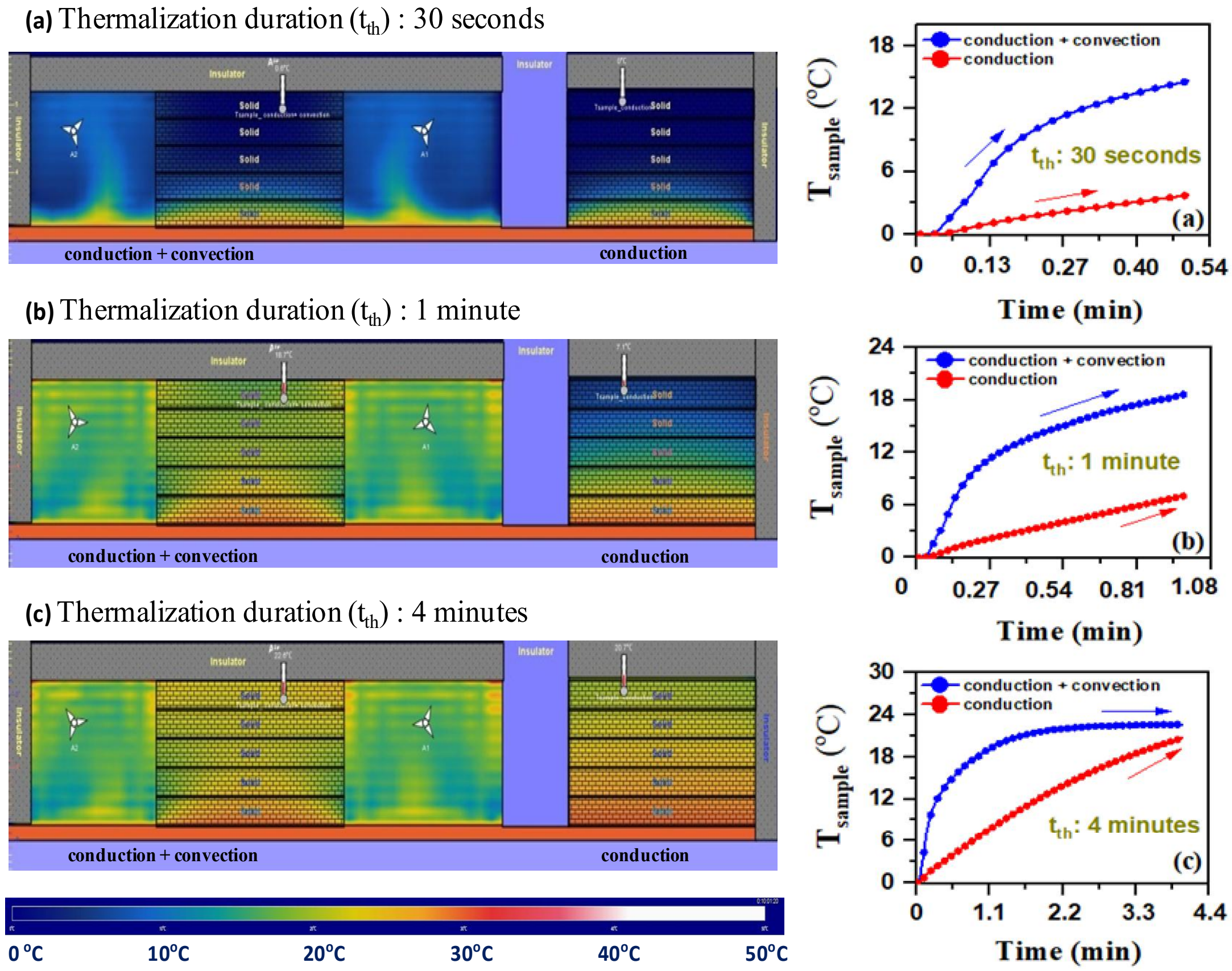

**Figure 6. Energy 2D simulation of thermal evolution under distinct heat-transfer conditions:** (a-c) Temperature distributions and corresponding sample-temperature evolution for thermalization durations ($t_{th}$): 30 seconds, 1 min, and 4 min, respectively. The gold-coated sample holder/insulating tape/Si/$SiO_2$/Mo and gold-coated sample puck/insulating tape/Si/$SiO_2$/NbN multilayer structures were assigned their respective thermal conductivities. An initial temperature gradient was established between the top and bottom surfaces, after which the system was allowed to approach thermal equilibrium. For each duration, the temperature distributions are compared for conduction-only heat transfer (right) and for combined conduction and convection (left). The combined conduction-and-convection condition results in a faster reduction in the temperature gradient and a faster approach to thermal equilibrium.

## Conclusion:

The combined transport measurements on Mo and NbN thin films establish that pronounced current-direction asymmetry can emerge from the interplay between current-induced Joule heating and the thermalization pathway of the measurement environment. In Mo, where the coexistence of metallic Mo and Mo-oxide phases provides a possible structural basis for weak-link transport, the strong suppression of the current-direction asymmetry under improved sample thermalization indicates that the observed diode-like response cannot be attributed solely to an intrinsic weak-link or superconducting-diode mechanism. This interpretation is further supported by NbN, where comparable hysteretic and asymmetric I-V characteristics are observed under conduction-dominated cooling despite the absence of the oxide-mediated weak-link structure proposed for Mo. The key physical point is that the superconducting-to-resistive transition can be strongly affected by electrothermal feedback. In the superconducting state, the dissipated power is very small because the resistance is negligible. As a resistive component develops near the switching condition, $P = I^2R$ increases, raising the sample temperature. The resulting finite thermal relaxation can alter both switching and retrapping conditions and produce hysteresis. Consequently, the experimentally measured switching current density (Jsw) should be distinguished from the intrinsic equilibrium switching current density (Jsw) whenever self-heating is appreciable. Because Joule heating depends on $I^2$, it is not intrinsically polarity dependent. The observed difference between the positive and negative branches should therefore not be interpreted as different Joule-heating powers for opposite current directions. Instead, it is consistent with the combined effects of current-induced heating, thermal history, the finite thermal relaxation time, and the effective electrical/thermal boundary conditions during the specified current sweep. The substantially more symmetric response obtained under improved thermalization provides experimental evidence that these electrothermal effects make a major contribution to the apparent diode-like transport. The thermal simulations further reproduce the distinct equilibration dynamics of the two cooling configurations, linking the transport response to the thermalization pathway rather than to nominal cooling capacity alone. These results demonstrate that measurement induced electrothermal effects can generate apparent diode-like superconducting transport even without an intentionally engineered nonreciprocal element. More broadly, they establish thermalization as a critical experimental parameter for identifying genuine superconducting diode effects and highlight the need to disentangle intrinsic nonreciprocal mechanisms from non-equilibrium thermal contributions in superconducting transport.

## Acknowledgements:

The authors acknowledge the Department of Atomic Energy (DAE), Government of India, for providing experimental support. One of the authors (Akhilesh Kumar), acknowledges the Senior Research Fellowship (SRF) from the Indira Gandhi Centre for Atomic Research (IGCAR), Kalpakkam, under the research programme of the Department of Atomic Energy, Government of India.

## References:

[1] Y. Tokura, N. Nagaosa, Nonreciprocal responses from non-centrosymmetric quantum materials, Nature Communications, 9 (2018) 3740.
[2] R. Wakatsuki, Y. Saito, S. Hoshino, Y.M. Itahashi, T. Ideue, M. Ezawa, Y. Iwasa, N. Nagaosa, Nonreciprocal charge transport in noncentrosymmetric superconductors, Science Advances, 3 e1602390.
[3] M. Nadeem, M.S. Fuhrer, X. Wang, The superconducting diode effect, Nature Reviews Physics, 5 (2023) 558-577.
[4] J. Ma, R. Zhan, X. Lin, Superconducting Diode Effects: Mechanisms, Materials and Applications, Advanced Physics Research, 4 (2025) 2400180.
[5] J.-K. Kim, K.-R. Jeon, P.K. Sivakumar, J. Jeon, C. Koerner, G. Woltersdorf, S.S.P. Parkin, Intrinsic supercurrent non-reciprocity coupled to the crystal structure of a van der Waals Josephson barrier, Nature Communications, 15 (2024) 1120.
[6] G.L.J.A. Rikken, E. Raupach, Observation of magneto-chiral dichroism, Nature, 390 (1997) 493-494.
[7] F. Liu, Y.M. Itahashi, S. Aoki, Y. Dong, Z. Wang, N. Ogawa, T. Ideue, Y. Iwasa, Superconducting diode effect under time-reversal symmetry, Science Advances, 10 eado1502.

[8] M. Trahms, L. Melischek, J.F. Steiner, B. Mahendru, I. Tamir, N. Bogdanoff, O. Peters, G. Reecht, C.B. Winkelmann, F. von Oppen, K.J. Franke, Diode effect in Josephson junctions with a single magnetic atom, Nature, 615 (2023) 628-633.
[9] S. Li, Z. Yang, X. Wang, K.P. Loh, X.R. Wang, Nonreciprocal electrical transport in emerging noncentrosymmetric systems: from hidden symmetry to functional devices, Reports on Progress in Physics, 89 (2026) 066501.
[10] S. Ilić, F.S. Bergeret, Theory of the Supercurrent Diode Effect in Rashba Superconductors with Arbitrary Disorder, Physical Review Letters, 128 (2022) 177001.
[11] S. Seki, Y. Okamura, K. Kondou, K. Shibata, M. Kubota, R. Takagi, F. Kagawa, M. Kawasaki, G. Tatara, Y. Otani, Y. Tokura, Magnetochiral nonreciprocity of volume spin wave propagation in chiral-lattice ferromagnets, Physical Review B, 93 (2016) 235131.
[12] F. Pop, P. Auban-Senzier, E. Canadell, G.L.J.A. Rikken, N. Avarvari, Electrical magnetochiral anisotropy in a bulk chiral molecular conductor, Nature Communications, 5 (2014) 3757.
[13] F. Ando, Y. Miyasaka, T. Li, J. Ishizuka, T. Arakawa, Y. Shiota, T. Moriyama, Y. Yanase, T. Ono, Observation of superconducting diode effect, Nature, 584 (2020) 373-376.
[14] F. Qin, W. Shi, T. Ideue, M. Yoshida, A. Zak, R. Tenne, T. Kikitsu, D. Inoue, D. Hashizume, Y. Iwasa, Superconductivity in a chiral nanotube, Nature Communications, 8 (2017) 14465.
[15] G. Hu, Y. Han, W. Yu, S. Lv, Y. Li, Z. Gong, H. Guo, K. Zhu, Z. Zhao, Q. Qi, G. Xian, L. Bao, X. Lin, J. Pan, S. Du, H. Yang, H.-J. Gao, Tunable zero-field superconducting diode effect in two-dimensional ferromagnetic/superconducting Fe3GeTe2/NbSe2 heterostructure, Nano Research, 18 (2025) 94907068.
[16] G. Carapella, P. Sabatino, G. Costabile, Asymmetry, bistability, and vortex dynamics in a finite-geometry ferromagnet-superconductor bilayer structure, Physical Review B, 81 (2010) 054503.
[17] A. Gutfreund, H. Matsuki, V. Plastovets, A. Noah, L. Gorzawski, N. Fridman, G. Yang, A. Buzdin, O. Millo, J.W.A. Robinson, Y. Anahory, Direct observation of a superconducting vortex diode, Nature Communications, 14 (2023) 1630.
[18] R. Lortz, Y. Wang, A. Demuer, P.H.M. Böttger, B. Bergk, G. Zwicknagl, Y. Nakazawa, J. Wosnitza, Calorimetric Evidence for a Fulde-Ferrell-Larkin-Ovchinnikov Superconducting State in the Layered Organic Superconductor kappa-$(BEDT\text{-}TTF)_2Cu(NCS)_2$, Physical Review Letters, 99 (2007) 187002.
[19] S. Reinhardt, T. Ascherl, A. Costa, J. Berger, S. Gronin, G.C. Gardner, T. Lindemann, M.J. Manfra, J. Fabian, D. Kochan, C. Strunk, N. Paradiso, Link between supercurrent diode and anomalous Josephson effect revealed by gate-controlled interferometry, Nature Communications, 15 (2024) 4413.
[20] P.J.W. Moll, V.B. Geshkenbein, Evolution of superconducting diodes, Nature Physics, 19 (2023) 1379-1380.
[21] C. Baumgartner, L. Fuchs, A. Costa, S. Reinhardt, S. Gronin, G.C. Gardner, T. Lindemann, M.J. Manfra, P.E. Faria Junior, D. Kochan, J. Fabian, N. Paradiso, C. Strunk, Supercurrent rectification and magnetochiral effects in symmetric Josephson junctions, Nature Nanotechnology, 17 (2022) 39-44.
[22] B. Pal, A. Chakraborty, P.K. Sivakumar, M. Davydova, A.K. Gopi, A.K. Pandeya, J.A. Krieger, Y. Zhang, M. Date, S. Ju, N. Yuan, N.B.M. Schröter, L. Fu, S.S.P. Parkin, Josephson diode effect from Cooper pair momentum in a topological semimetal, Nature Physics, 18 (2022) 1228-1233.
[23] K.-R. Jeon, J.-K. Kim, J. Yoon, J.-C. Jeon, H. Han, A. Cottet, T. Kontos, S.S.P. Parkin, Zero-field polarity-reversible Josephson supercurrent diodes enabled by a proximity-magnetized Pt barrier, Nature Materials, 21 (2022) 1008-1013.
[24] J.J. He, Y. Tanaka, N. Nagaosa, The supercurrent diode effect and nonreciprocal paraconductivity due to the chiral structure of nanotubes, Nature Communications, 14 (2023) 3330.
[25] M.S. Anwar, T. Nakamura, R. Ishiguro, S. Arif, J.W.A. Robinson, S. Yonezawa, M. Sigrist, Y. Maeno, Spontaneous superconducting diode effect in non-magnetic $Nb/Ru/Sr_2RuO_4$ topological junctions, Communications Physics, 6 (2023) 290.

[26] R.-P. Yu, J.-X. Hu, Z.-T. Sun, Transverse superconducting diode without violation of parity- or time-reversal symmetry, Physical Review Applied, 26 (2026) L011002.
[27] J. Hu, C. Wu, X. Dai, Proposed Design of a Josephson Diode, Physical Review Letters, 99 (2007) 067004.
[28] J. Wang, Y. Jiang, J.J. Wang, J.-F. Liu, Efficient Josephson diode effect on a two-dimensional topological insulator with asymmetric magnetization, Physical Review B, 109 (2024) 075412.
[29] S.-L. Wu, Z.-H. Ren, P. Zhu, H.-C. Zhang, X.-T. Di, C. Wang, C. Li, Z. Wang, C.-Z. Li, Z.-M. Liao, Tunable Josephson diode effect mediated by topological surface states in $BiSbTeSe_2$ Josephson junctions, Physical Review B, 113 (2026) 155435.
[30] J. Ma, H. Wang, W. Zhuo, B. Lei, S. Wang, W. Wang, X.-Y. Chen, Z.-Y. Wang, B. Ge, Z. Wang, J. Tao, K. Jiang, Z. Xiang, X.-H. Chen, Field-free Josephson diode effect in NbSe2 van der Waals junction, Communications Physics, 8 (2025) 125.
[31] H. Wu, X. Zhou, J. Li, X. Li, B. Li, W. Fei, J. Zhou, J. Yin, W. Guo, Ultrathin Molybdenum Dioxide Nanosheets as Uniform and Reusable Surface-Enhanced Raman Spectroscopy Substrates with High Sensitivity, Small, 14 (2018) 1802276.
[32] A.V. Gurevich, R.G. Mints, Self-heating in normal metals and superconductors, Reviews of Modern Physics, 59 (1987) 941-999.
[33] J. Viña, M.T. González, M. Ruibal, S.R. Currás, J.A. Veira, J. Maza, F. Vidal, Self-heating effects on the transition to a highly dissipative state at high current density in superconducting $YBa_2Cu_3O_{7-\delta}$ thin films, Physical Review B, 68 (2003) 224506.
[34] C. Xie, Interactive Heat Transfer Simulations for Everyone, The Physics Teacher, 50 (2012) 237-240.
[35] H.R. Shanks, P.D. Maycock, P.H. Sidles, G.C. Danielson, Thermal Conductivity of Silicon from 300 to 1400°K, Physical Review, 130 (1963) 1743-1748.
[36] D.-M. Zhu, Thermal conductivity of densified $\alpha$-SiO2 from 15 K to room temperature, Physical Review B, 50 (1994) 6053-6056.
[37] S.H. Firoz, T. Yagi, N. Taketoshi, K. Ishikawa, T. Baba, Direct observation of thermal energy transfer across the thin metal film on silicon substrates by a rear heating–front detection thermoreflectance technique, Measurement Science and Technology, 22 (2011) 024012.